\documentclass[]{pasj01}
\draft

\usepackage{color,ulem}

\begin{document} 
\Received{}%{yyyy/mm/dd}
\Accepted{}%{yyyy/mm/dd}
%\Published{yyyy/mm/dd}

\title{Suzaku Observation of Diffuse X-Ray Emission from a Southwest Region of the Carina Nebula}

\author{Yuichiro \textsc{Ezoe},\altaffilmark{1}}
\altaffiltext{1}{Department of Physics, Tokyo Metropolitan University, 1-1 Minami-Osawa, Hachioji, Tokyo 192-0397, JAPAN}

\author{Kenji \textsc{Hamaguchi},\altaffilmark{2,3}}
\altaffiltext{2}{CRESST II and X-ray Astrophysics laboratory NASA GSFC, Greenbelt, MD20771, USA}
\altaffiltext{3}{Department of Physics, University of Maryland, Baltimore Country, 1000 Hiltop Circle, Baltimore MD21250, USA}

\author{Aoto \textsc{Fukushima},\altaffilmark{1}}

\author{Tomohiro \textsc{Ogawa},\altaffilmark{1}}

\author{Takaya \textsc{Ohashi},\altaffilmark{1}}

\KeyWords{ISM: abundances --- ISM: individual objects (Carina Nebula) --- stars: winds, outflows --- ISM: supernova remnants --- X-rays: ISM}

\maketitle

\begin{abstract}
A southwest region of the Carina nebula was observed with the {\it Suzaku} observatory for 47 ks in 2010 December.
This region shows distinctively soft X-ray emission in the {\it Chandra} campaign observations.
{\it Suzaku} clearly detects the diffuse emission
above known foreground and background components between 0.4$-$5~keV
at the surface brightness of 3.3$\times10^{-14}$ erg s$^{-1}$ arcmin$^{-2}$.
The spectrum requires two plasma emission components with {\it kT} $\sim$0.2 and 0.5 keV, 
which suffer interstellar absorption of $N_{\rm H}$ $\sim$1.9$\times10^{21}$ cm$^{-2}$.
Multiple absorption models assuming two temperature plasmas at ionization equilibrium or
non-equilibrium are tested but there is no significant difference in terms of $\chi^2$/d.o.f..
These plasma temperatures are similar to those of the central and eastern parts of the Carina nebula measured 
in earlier {\it Suzaku} observations, but the surface brightness of the hot component is significantly lower than those of the other regions.
This means that these two plasma components are physically separated and have different origins.
The elemental abundances of O, Ne and Mg with respect to Fe favor that the diffuse plasma originates from 
core-collapsed supernovae or massive stellar winds.
\end{abstract}

%%%%%%%%%%%%
\section{Introduction}
%%%%%%%%%%%%

With significant advances in X-ray observing techniques, diffuse X-ray emission has been detected from
many massive star forming regions. Nearby ($\lesssim$2 kpc) samples include RCW 38 \citep{wol02}, 
the Rosette nebula \citep{tow03}, M17 \citep{tow03, hyo08}, NGC 6334 \citep{ezo06a}, NGC 2024 \citep{ezo06b}, 
the Carina nebula \citep{ham07, ezo09, tow11a}, the Orion nebula \citep{gue08} and 
the Cyg OB2 association \citep{alb18}.
Their spectra are characterized by multiple-temperature thermal plasma emission
with typical plasma temperatures below 1 keV and luminosities of $10^{32-35}$ erg s$^{-1}$ and
some of them (e.g., RCW38, NGC 6334, NGC 2024)
also show a hint of a power-law spectrum.
Among these nearby star forming regions, the Carina nebula has the largest diffuse X-ray luminosity at $\sim$10$^{35}$ erg s$^{-1}$.

Two mechanisms are considered for the origin of these diffuse plasmas, 
which drive fast powerful winds and produce X-ray emitting hot plasma bubbles \citep{cas75,wea77}.
Alternatively, some massive stars born in those nebulae could have exploded as supernovae (SNe), whose remnants
may be deformed heavily by dense circumstellar medium and and/or merged with the other SN remnants.

Key diagnostics to understand the origin
would be properties of the thermal plasma including temperatures, spatial distributions
and elemental abundances. For example, 
the plasma abundance should be as abundant as solar value if 
it is heated up by the stellar wind \citep{cun94, daf04}, while the plasma should
be overabundant in iron if it originates from
a Type Ia SN \citep{tsu95}.
This plasma diagnostics is especially effective for cool plasmas ($\sim$1 MK)
suffering little absorption ($N_{\rm H}\lesssim$10$^{21}$ cm$^2$), 
which emit X-ray emission from highly striped ions such as carbon, oxygen,
iron, magnesium, silicon and sulfur. 

\begin{figure}[htbp]
	\centerline{
		\includegraphics[width=0.725\textwidth]{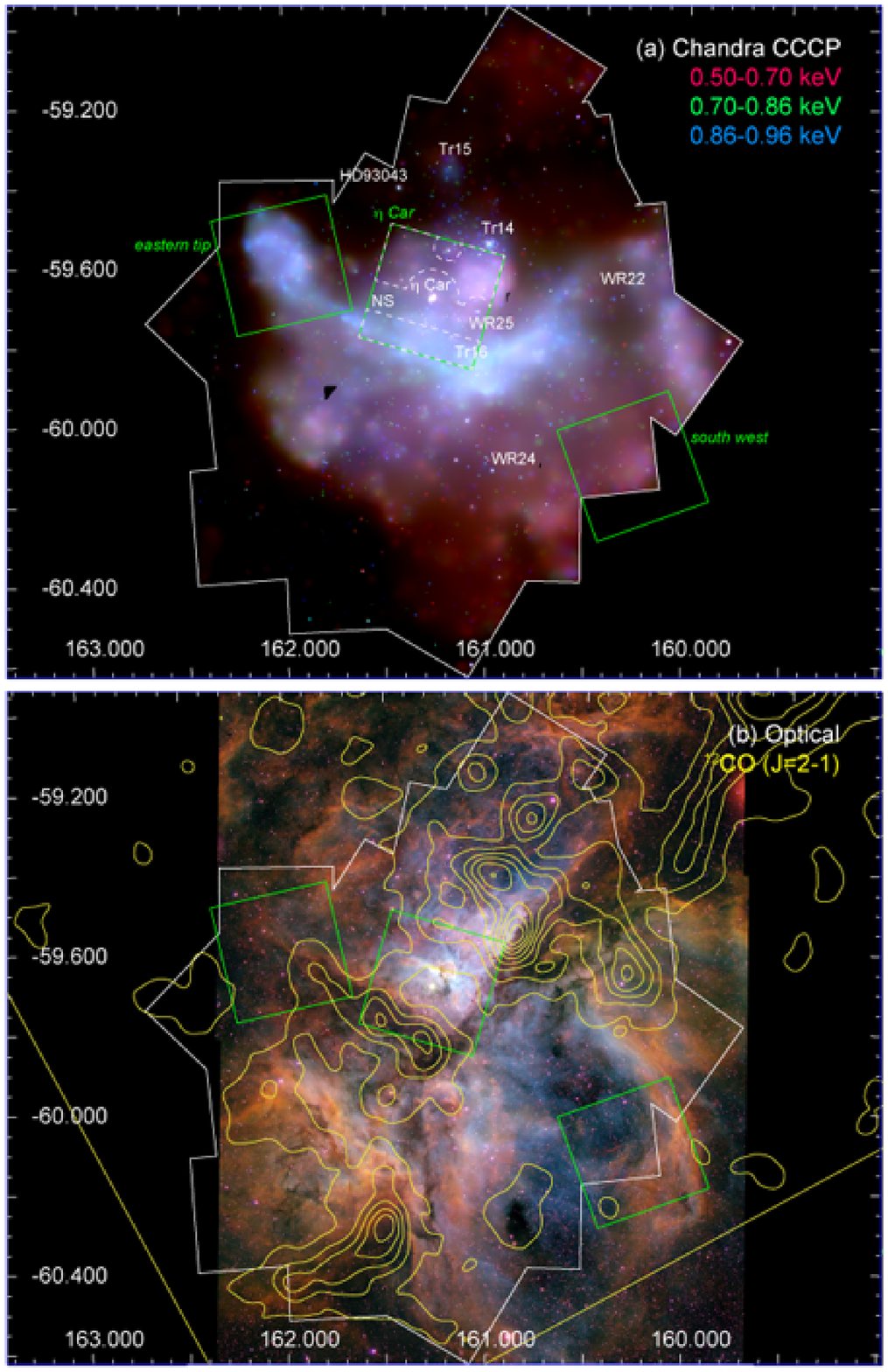}
	}
	\caption{(a) Color-coded Chandra X-ray (red: 0.5--0.7 keV, green: 0.7--0.86 keV, blue: 0.86--0.96 keV) image of the Carina nebula provided 
	from the CCCP \citep{tow11a}.
	The white labels indicate major massive stellar clusters, massive O main-sequence/Wolf-Rayet stars and the middle-age neutron star (NS) \citep{ham09}. 
	(b) Optical emission line (blue: oxygen, green: hydrogen, red: sulfur) image
	of the Carina nebula (Nathan Smith, University of Minnesota/NOAO/AURA/NSF, See https://www.noao.edu/image\_gallery/html/im0667.html). 
	The blue/red structures tend to be emitted from hotter/cooler ionized gases.
	The yellow lines show a $^{12}$CO (J$=$2-1) contour map and the mapping boundary \citep{yon05}.
	(a\&b) The three green boxes show the FOV of the {\it Suzaku} XIS observations used for the diffuse emission ---
	the eastern-tip, $\eta$ Car, and southwest regions from left.
	The white dash lines in the $\eta$ Car region show subdivided areas (north, south) for analysis in \citet{ham07}.
	The solid white line shows the CCCP survey area. Coordinates are J2000. 
	A 5$^\prime$ angular scale corresponds to 3.4 pc at $d$ = 2.3 kpc.
	}
\label{fig1}
\end{figure}

The Carina nebula is one of the most active massive star forming regions in our Galaxy. It comprises
8 open clusters with at least 66 O stars, 3 Wolf-Rayet stars, and the luminous blue variable $\eta$ Car \citep{smi08}.
It is located at $\sim$2.3 kpc, closer than its rivals such as NGC 3603 and W49 ($\gtrsim7$ kpc).
\citet{sew79} first suggested extended X-ray emission from the nebula using the {\it Einstein} observatory.
The Carina Complex Project (CCCP) conducted a wide field ($\sim$1.4 deg$^2$) survey of the Carina nebula
with {\it Chandra} ACIS-I \citep{tow11a} and 
detected soft extended X-ray emission as well as $>$14000 X-ray point sources (figure \ref{fig1}). \citep{tow11a,tow11b,tow11c}. 
From a study of the X-ray point sources, they concluded that the majority of the extended X-ray emission does not originate from unresolved faint point sources but from truly diffuse plasma extended over the interstellar space. 
\citet{tow11b} also suggested that
a spectrum stacked from regions with enhanced iron lines may
show emission lines at 0.56, 0.76, and 1.85 keV, 
possibly from charge exchanges of
highly ionized oxygen atoms (OVII and OVIII) in 
hot plasmas that interact with surrounding cold interstellar materials.

The Suzaku observatory has provided important clues on the property of the diffuse emission thanks to the good sensitivity 
and energy resolution in the soft band.
\citet{ham07} studied diffuse X-ray emission from north and south regions of $\eta$ Car.
Both spectra can be reproduced 
by emission models of two temperature plasma at $kT\sim0.2$ and $\sim$0.6 keV,
but they show a distinct difference at $\sim$1 keV, which requires spatial variation 
by a factor of 4.

\citet{ezo09} studied diffuse X-ray emission from the eastern tip region of the nebula 
using both {\it Suzaku} and {\it XMM-Newton} data.
The spectrum was similar to that of the $\eta$ Carinae south region. 
Since this region does not contain any OB stars nor known compact objects, 
the diffuse plasma was originally thermalized inside the cluster and escaped out to 
this region and/or it was thermalized by collision of the ambient cold medium with high 
speed winds flown from the cluster center with OB stars. 
These results suggest that the diffuse X-ray plasmas originate from 
either stellar wind bubbles or SNe.
The supernova hypothesis is supported by a discovery of a middle age neutron star
located at $\sim8'$ southeast of $\eta$ Car \citep{ham09}.

This paper studies diffuse X-ray emission from a southwest region of the Carina nebula using the X-ray data 
obtained with the Suzaku observatory in 2010.
This is the third and final observation of the Carina nebula with {\it Suzaku},
whose mission ended in 2015 August.
This observing field is distinctively red in the Chandra CCCP X-ray true color image, i.e., very soft X-ray emission.
This study aims to understand the cause of this redness with the {\it Suzaku}'s good sensitivity and energy resolution
in the soft X-ray band.

%%%%%%%%%%%%
\section{Observation}
%%%%%%%%%%%%

The X-ray astronomy satellite {\it Suzaku} \citep{mit07} observed a southwest region of the Carina nebula on 2010 December 12
(observation ID : 505075010). 
At the time of this observation, {\it Suzaku} operated two types of detectors,
the X-ray Imaging Spectrometer (XIS: \cite{koy07}) and the Hard X-ray Detector (HXD: \cite{tak07}).
We use only the XIS, which are sensitive to the soft X-ray emission from the diffuse plasma.
Three XIS cameras were available during the observation; two of them (XIS0, 3) have front-illuminated (FI) CCDs 
and one (XIS1) has a back-illuminated (BI) CCD.
The XIS2 camera was lost in 2006 due to a micrometeoroid impact\footnote{http://www.astro.isas.ac.jp/suzaku/doc/suzaku\_td/}.
The {\it Suzaku} XISs have lower particle background, higher effective area and better energy resolution for soft extended sources
than instruments onboard earlier or ongoing X-ray observatories.
They are, therefore, suited best for observing diffuse X-ray emission from the Carina nebula.

We download the final processing data from the HEASARC archive (processing version 3.0.22.43)
\footnote{See http://www.astro.isas.jaxa.jp/suzaku/analysis/ for more details on the processing version.}
and analyze the standard cleaned event data using the HEAsoft analysis package version 6.21.
We exclude events detected at the flickering pixels that are identified by the instrument team
from non X-ray background (NXB) data throughout the {\it Suzaku} mission\footnote{http://www.astro.isas.jaxa.jp/suzaku/analysis/xis/nxb\_new2/}.
This procedure completely excludes noise events in the low energy band, which are significant in the later {\it Suzaku} observing cycle.
We produce background images and spectra with these NXB data using {\tt xisnxbgen}.
The total exposure time is 47.0 ks.

%%%%%%%%%%%%%%%
\section{Spatial distribution}
%%%%%%%%%%%%%%%

Figure~\ref{fig1} shows a true color {\it Chandra} X-ray image of the Carina nebula provided by the CCCP project \citep{tow11a}
and an optical emission line image with $^{12}$CO contours.
In figure~\ref{fig1} (a), the surface brightness is the highest near the image center, offset from the active star 
forming clusters, Trumpler 14 and 16 to the south.
In this X-ray image, the two blue streaks
emanate from this brightest region to the east-northeast and northwest directions.
The area between eta Carinae and Trumpler 14 also has high surface X-ray brightness, 
but the X-ray color is redder than those of the two streaks.
%The surface brightness is also high between $\eta$ Carinae and Trumpler 14, which is redder than the two streaks.
The surface brightness is lower and the color is redder toward the southern outskirts.

The first {\it Suzaku} study of the diffuse X-ray emission \citep{ham07} covered the high surface brightness regions.
The north and south regions were analyzed separately as their spectra show distinct differences at $\sim$1 keV.
The second study \citep{ezo09} covered the crescent-like plasma structure at the tip of the east-northeast blue streak.
The third {\it Suzaku} study described in this paper focuses on a red blob near a southwest edge of the CCCP survey field.
The rightmost green box in Figure~\ref{fig1} (a) shows the XIS FOV of this observation,
a half of which is outside of the CCCP survey field.

\begin{figure}[t]
	\centerline{
		\includegraphics[width=0.7\textwidth]{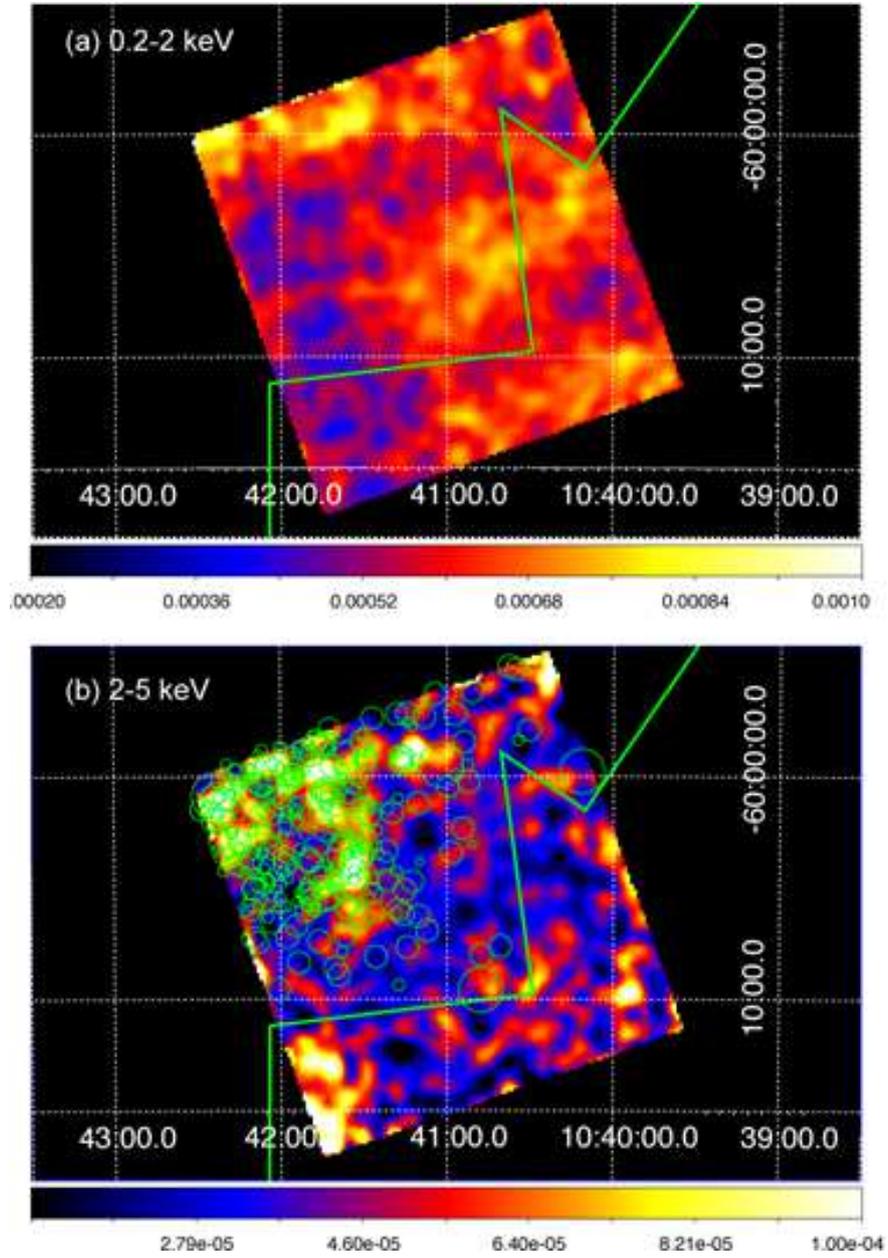}
	}
	\smallskip
	\caption{(a) {\it Suzaku} XIS1 
	images of the southwest region of the Carina nebula in the (a) 0.2--2 keV and (b) 2--5 keV bands. 
	These images are %binned by 8 pixels (default in the Suzaku data analysis) and 
	smoothed by a 	2-dimensional Gaussian function with $\sigma=$5 pixel.
	The green solid lines 
	indicate the boundary of the CCCP image (the upper left side was imaged.).
%	The white solid lines show subdivided regions for the {\it Suzaku} spectral analysis.
	The green circles in the lower panel show positions of the Chandra point sources \citep{tow11a}.
	The radius of each circle expresses the 0.5--8 keV flux of the corresponding source in \citet{bro11};
	$r=$10\" ~for $< $F$_{\rm X}$ 10$^{-16}$, 15\" ~for 10$^{-16}\sim$10$^{-15}$,
	30\" ~for 10$^{-15}\sim$10$^{-14}$, and 60\" ~for $>$10$^{-14}$ erg s$^{-1}$ cm$^{-2}$.
	The coordinates system is J2000.}
\label{fig2}
\end{figure}

Figure~\ref{fig2} shows soft (0.2$-$2 keV) and hard (2$-$5 keV) band images of the {\it Suzaku} XIS observation 
of the southwest region.
For each image, the NXB contribution is estimated from night Earth observations with {\tt xisnxbgen} and 
subtracted from the raw image.
Then, the vignetting effect is corrected with a flat field map at 1.49 keV, which is generated with {\tt xissim}\footnote{https://heasarc.gsfc.nasa.gov/docs/suzaku/analysis/expomap.html\#trim}.
The soft band image shows a small flux enhancement around the center,
which is seen as a blob in the CCCP image (Figure~\ref{fig1}a).
The hard band image shows an enhancement in the upper left 
where the CCCP project detected multiple point sources \citep{tow11a}.

%%%%%%%%%%%%
\section{Spectral Analysis}
%%%%%%%%%%%%

\begin{figure}[p]
	\centerline{
		\includegraphics[width=0.55\textwidth]{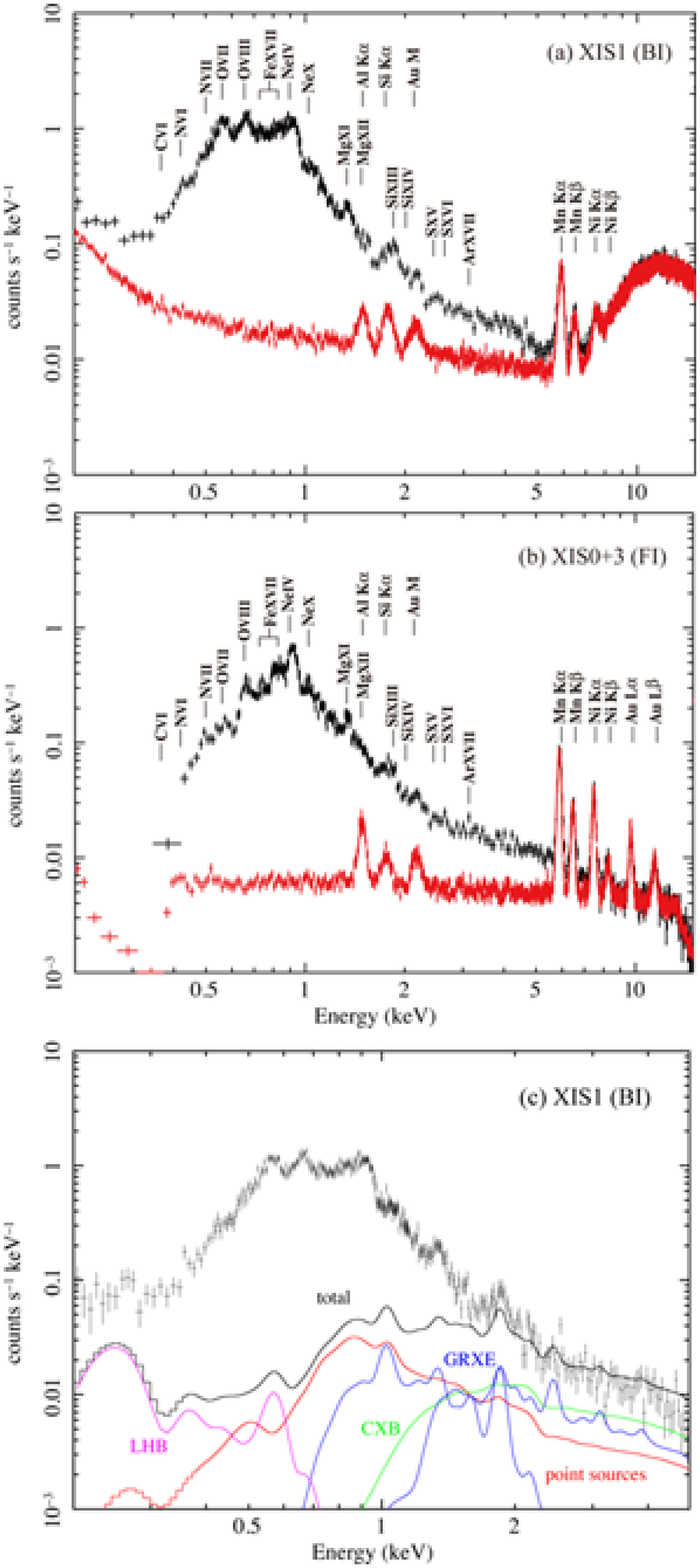}
	}
	\smallskip
	\caption{{\it Suzaku} BI (a) and FI (b) spectra of the whole XIS FOV (black). The  red spectra show 
	show NXB contribution, estimated from the night Earth observations.
	The labels denote emission lines detected in the source and/or background spectra.
	(c) Background-subtracted BI spectrum of the whole XIS FOV. The solid lines indicate estimated 
	contamination contributions.}
\label{fig3}
\end{figure}

We extract XIS FI (XIS0$+$3) and BI (XIS1) spectra of the whole FOV (figure~\ref{fig3})
and generate the corresponding NXB spectra from the night Earth data using {\tt xisnxbgen}.
The source and background spectra of each sensor match well above $\sim$10 keV where the XIS have no sensitivity in X-rays,
confirming that the NXB spectra are reproduced correctly.
The 0.2--5 keV count rates before and after subtracting the NXB are 0.793$\pm$0.004 and 0.723$\pm$0.004~counts s$^{-1}$ for BI
and 0.333$\pm$0.002 and 0.305$\pm$0.002 counts s$^{-1}$ for one FI
where the errors are 1$\sigma$.

The NXB subtracted spectra (figure~\ref{fig4} for BI) show emission lines from highly ionized ions of various elements. 
such as O{\sc vii}, O{\sc viii}, Ne{\sc iv}, Ne{\sc x}, Mg{\sc xi}, and S{\sc xv}.
\citet{tow11b} reported a marginal emission line at 1.3 keV from the Carina SW region, but
the {\it Suzaku} spectra show no emission line at this energy.

%%%%%%%%%%%%
\subsection{Contamination from astrophysical backgrounds and point sources}
%%%%%%%%%%%%

The extracted spectra still include background or foreground sources 
--- local hot bubble (LHB), galactic ridge X-ray emission (GRXE), cosmic X-ray background (CXB), and 
point sources mostly unresolved with {\it Suzaku}.
We estimate contributions of LHB, GRXE, and CXB by following the method in \citet{ezo09} and \citet{ham07}.

\begin{figure}[tbp]
	\centerline{
		\includegraphics[width=0.8\textwidth]{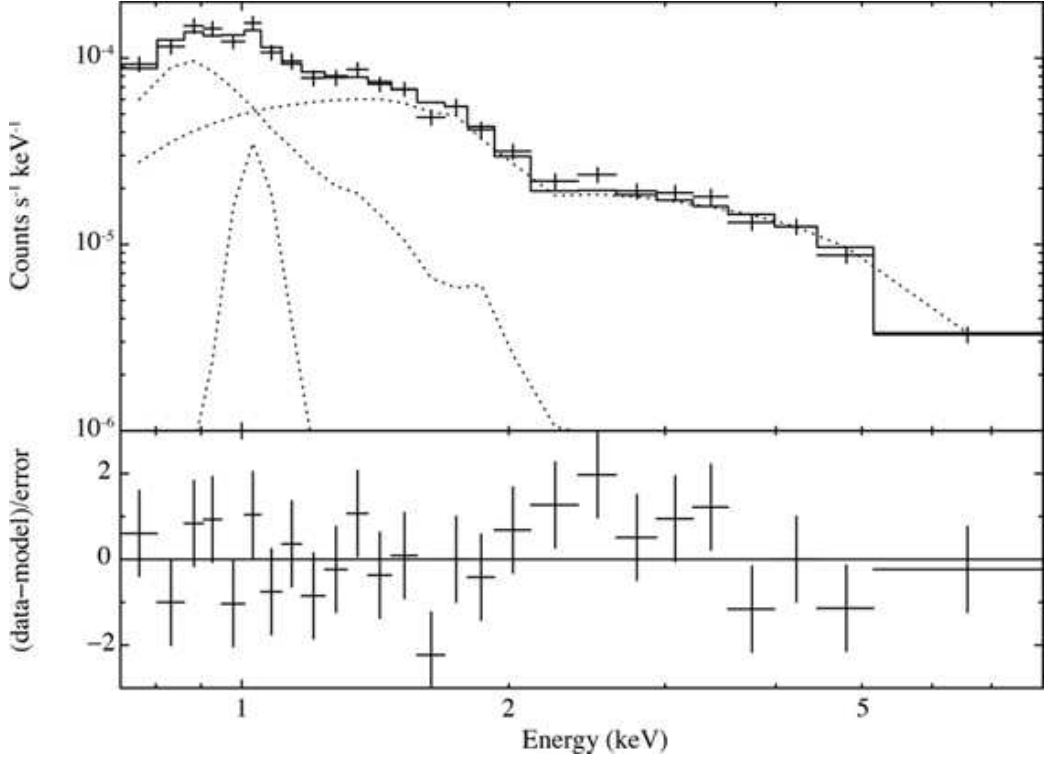}
	}
	\smallskip
	\caption{Average {\it Chandra} ACIS-I spectrum of the CCCP point sources. The dotted lines show the individual components of the best-fit model.}
\label{fig4}
\end{figure}

We then estimate contribution of X-ray point sources inside the {\it Suzaku} FOV using the CCCP product.
{\it Chandra} observed this field, labelled as Supperbubble3 (SB3),
twice in 2008 May 24 and 31 (ObsID 9498 and 9859)  for 31.9 ks and 26.8 ks, respectively.
The CCCP catalogue \citep{bro11} lists 400 point sources inside the {\it Suzaku} FOV,
whose average flux was 2.7$\times10^{-15}$ erg s$^{-1}$ cm$^{-2}$ between 0.2--5 keV.
A combined {\it Chandra} spectrum of these point sources provided by the CCCP team 
(Private Comm.: Broos, P.) extends up to $\sim$8~keV (figure \ref{fig4}), suggesting a very hot plasma or flat power-law emission. 
The spectrum also shows line-like structures below $\sim$1~keV, suggesting 
significant contribution of cool thermal plasma emission from low-mass young stars in the field.
We fit the spectrum by a model with a single-temperature plasma emission component ({\tt apec}) plus a power-law component,
both of which suffer interstellar absorption to the Carina nebula. %% ($N_{\rm H}$ $\sim$1$\times$10$^{22}$ cm$^{-2}$).
The elemental abundance of the thermal component is fixed at 0.3 solar, the typical 
abundance of stellar X-ray plasma (e.g., \cite{get05}).
We use the corresponding response (rmf) and ancillary response (arf), weighted for the source fluxes.

The best-fit of the {\tt apec} $+$ {\tt power-law} model has a residual at $\sim$1~keV, not acceptable with $\chi^2$/d.o.f. at 1.75. 
Adding a narrow Gaussian line significantly improves the fit in an F-test ($\sim$98 \%) with $\chi^2$/d.o.f. at 1.28.
The best-fit centroid energy is consistent with a Ne X Ly$_\alpha$ line (1022 eV).
Assuming non-equilibrium plasma emission model ({\tt nei}) also improves 
a fit with similar $N_{\rm H}$ and $kT$ and an ionization parameter at $\tau =1.3^{+3.3}_{-1.0}$ s cm$^{-3}$. 
However, the $\chi^2$/d.o.f. is 1.41, worse than the model with a Gaussian line.
Therefore, we employ the second {\tt apec} $+$ {\tt Gaussian} $+$ {\tt power-law} model to reproduce the point source flux.
The best-fit model is shown in table \ref{tbl1}.
The total X-ray fluxes in the 0.2--2 keV and 2--5 keV bands are 7.1$\times10^{-13}$ erg s$^{-1}$ cm$^{-2}$
and 3.7$\times10^{-13}$ erg s$^{-1}$ cm$^{-2}$, respectively.

\begin{table}[p]

\caption{The best-fit result of the average {\it Chandra} point source spectrum. The model assumes {\tt tbabs (apec + Gaussian + power-law)}.} 
\label{tbl1}

\begin{center}

	\begin{tabular}{cc} \hline
	
	Interstellar absorption\\
	$N_{\rm H}$ (10$^{22}$ cm$^2$)	 &  0.053 ($<$0.252) \\
	\hline
	
	Plasma component$^*$ \\
	$kT$ (keV) 	      &  0.74$\pm0.10$ \\
	$Z$   (solar)            & 0.3 (fixed) \\
	Normalization  &  3.5$^{+4.6}_{-1.2}\times10^{-7}$\\
	\hline

	Gaussian component\\
	$E_{\rm center}$ & 1.04$^{+0.03}_{-0.02}$\\
	$\sigma$          &  0 (fixed)\\
	Normalization (photons s$^{-1}$ cm$^{-2}$) & 1.5$^{+0.3}_{-0.1}\times10^{-8}$\\ 
	\hline
		
	Power law component$^\dagger$ \\
	$\Gamma$     & 1.0$\pm0.1$\\
	Normalization (keV$^{-1}$ s$^{-1}$ cm$^{-2}$ at 1 keV) & 2.1$^{+0.3}_{-0.2}\times10^{-7}$\\
	\hline

	X-ray flux, luminosity$^\S$\\
%	$F_{\rm X}$ in 0.3--5 keV (erg s$^{-1}$ cm$^{-2}$) & 1.7$\times 10^{-15}$\\
%	$L_{\rm X}$ in 0.3--5 keV (erg s$^{-1}$) & 2.7$\times 10^{29}$\\
	
	$F_{\rm X}$ in 0.3--7 keV (erg s$^{-1}$ cm$^{-2}$) & 2.3$\times 10^{-15}$\\
	$L_{\rm X}$ in 0.3--7 keV (erg s$^{-1}$) & 3.6$\times 10^{29}$\\
	
	\hline
	
	$\chi^2$/d.o.f  & 1.28 \\
	d.o.f  	       & 19 \\
	\hline

	\end{tabular}
	
\end{center}
	$^*$ The normalization is defined as $10^{-14}/(4\pi D^2) EM$, where $D$ is the distance to the Carina nebula in cm 
	and $EM$ is the plasma emission measure in cm$^{-3}$. The elemental abundances are given relative to the solar photospheric values 
	measured by \citet{and89}.\\
	$^\S$ A distance of 2.3 kpc is assumed. The total flux and luminosity of all point sources are $\times$400 of these values.
\end{table}

Figure \ref{fig3} (c) plots these contamination components on the XIS1 spectrum.
The X-ray collecting area changes with the position  and  spatial distribution of the source on the detector.
The collecting area for each emission component is 
calculated with {\tt xissimarfgen} and the result is stored in an arf file.
For the {\it Chandra} point sources,
we assume a {\it Chandra} ACIS image between 0.5--7 keV
taken on 2008 May 24 (ObsID 9498), which has a longer exposure of the two Supperbubble3 observations,
as the point source distribution.
For the other components, we do not expect strong spatial variation in arcmin scales, and thus
we assume uniform surface brightness within 20$'$ from the detector center. 
We reproduce the XIS1 spectra of the contamination components using these arfs.

The total of the contamination spectra matches moderately well with the observed spectrum above 2~keV.
This means that the high energy spectrum mostly originates from GRXE and CXB.
The small discrepancy can be explained by
i) some point sources are active galactic nuclei, whose contribution is also counted in the CXB,
ii) {\it Chandra} detected point sources only from a half of the XIS FOV, and
iii) the CXB fluctuates $\sim$5\% (1 $\sigma$) in the sky \citep{kus02}.

Below $\sim$0.4~keV, the observed spectrum is a factor of $\sim$3 higher
than the total spectrum of the known source, mostly the LHB.
This excess emission should originate from a foreground source
as soft X-ray emission below 0.4~keV 
from sources more distant than the Carina nebula is totally obscured by interstellar absorption.
An obvious candidate is the solar wind charge exchange (SWCX)
--- an interaction of solar winds with material around the Earth (e.g., \cite{sno04,fuj09,ezo10,ezo11,ish13}) ---
which produces multiple emission lines below 1~keV.
However, this component correlates with the solar wind intensity and
varies on typical timescales of 1$-$10 ksec,
but the XIS1 light curve between 0.2--0.6 keV does not show any significant variation.
In addition, average proton and ion (C$^{6+}$ and O$^{7+}$) fluxes of the solar wind,
measured with the solar wind monitoring satellites WIND\footnote{http://web.mit.edu/space/www/wind/wind\_data.html} and ACE\footnote{http://www.srl.caltech.edu/ACE/ASC/level2/index.html},
are $\sim$5 times smaller around this observation than those during earlier SWCX events observed with {\it Suzaku}.
These results do not suggest the contamination of significant SWCX emission during this observation.
This soft excess may originate from a local enhancement of the LHB emission.

%%%%%%%%%%%%
\subsection{Spectral fitting}
%%%%%%%%%%%%

The observed spectrum between 0.4$-$1.5 keV is 2$-$30 times higher than the sum of these contamination components
in count rates
and therefore should originate mostly from truly diffuse X-ray plasma in the Carina nebula.
We simultaneously fit the XIS BI and FI spectra to constrain the spectral parameters.
For the truly diffuse component, we assume an absorbed ({\tt phabs}) two temperature collisional equilibrium plasma ({\tt apec}) model, 
which is also used for earlier Suzaku studies of the Carina nebula. 
For the cosmic background and foreground emission,
we add a power-law model to reproduce the hard spectrum above 2 keV and
a thermal ({\tt apec}) model to reproduce the soft spectrum below 0.4 keV.
For the instrumental response, we use {\tt arf} files that assume uniformly extended emission.

\begin{figure}[htbp]
	\centerline{
		\includegraphics[width=0.8\textwidth]{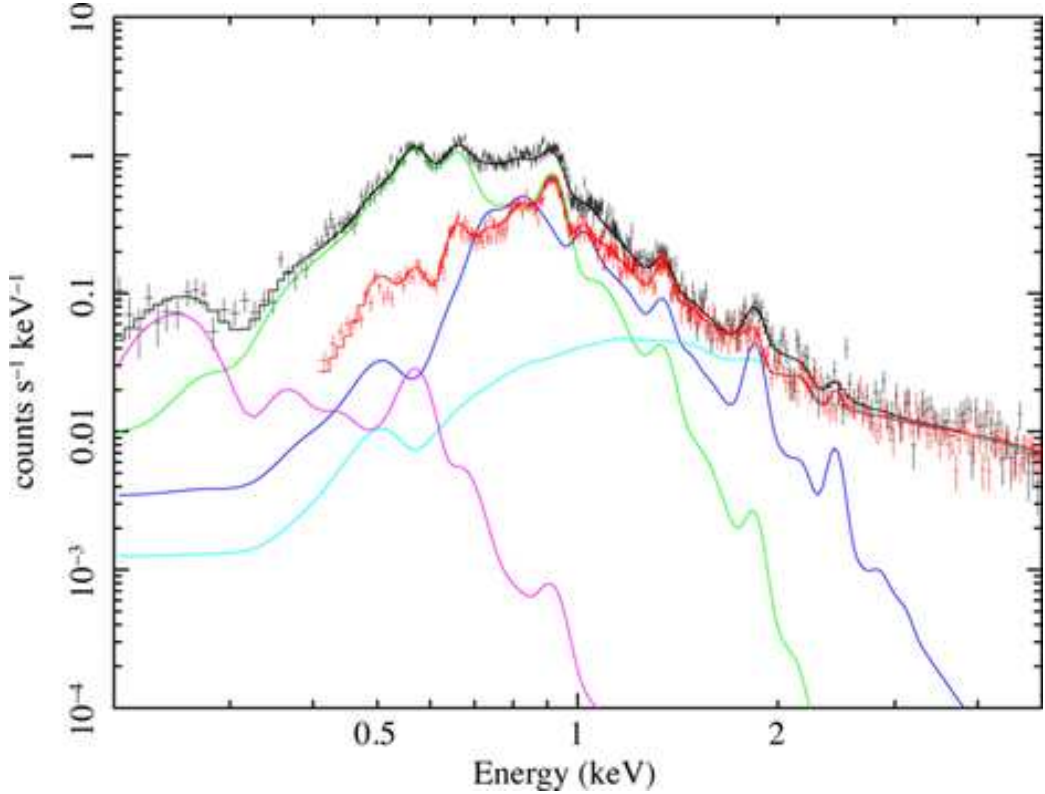}
	}
	\smallskip
	\caption{The NXB subtracted spectra of the whole XIS FOV. The black and red solid lines show the best-fit model for the BI and FI spectra.
	The solid lines in the other colors show the best-fit of the individual components for the XIS1 spectrum.}
\label{fig5}
\end{figure}

Elemental abundances of O, Ne, Mg, Si, S, and Fe, whose emission lines are seen clearly in the spectra (figure \ref{fig3}),
are allowed to vary independently, while those of the other elements are fixed at 0.3 solar values.
Figure \ref{fig5} and Table \ref{tbl:fit1} show the best-fit result. 
We calculate the X-ray fluxes and luminosities in 0.4--5 keV where the diffuse X-ray emission dominates. 
Also as a reference, we also show the X-ray fluxes between 0.5--7 keV, which is often used for Chandra studies 
of diffuse X-ray emission from star forming regions.

The hydrogen column density, $N_{\rm H} \sim$1.9$\times10^{21}$ cm$^{-2}$, is
similar to those of the other Carina regions (1.2$-$2.4$\times10^{21}$ cm$^{-2}$, \cite{ham07,ezo09}).
The plasma temperatures, $\sim$0.17 and 0.50~keV, are only slightly lower than those of 
diffuse X-ray plasmas in the other Carina regions studied with {\it Suzaku} (0.20--0.25 keV and 0.54--0.60 keV, \cite{ham07,ezo09}).
This result suggests that the plasma are heated in a similar mechanism to those 
in the other Carina nebula regions.

%%% Table 

\begin{table}[p]
  \caption{The best-fit two-temperature plasma model for spectra of the whole region and subdivided regions.}

  \label{tbl:fit1}

  \begin{center}
    \begin{tabular}{lccccc}

\hline
Model$^{\rm a}$                               &  whole region  & inner & outer \\
\hline									       	        								   
									       	        								   
Two-temperature plasma component$^{\rm b}$\\				       	        			   

% 2019.06.27
% center2 (LHB fix), outer2								       	        								   
$N_{\rm H}$ (10$^{22}$ cm$^{-2}$)	& 0.189$\pm{0.002}$ & 0.169$^{+0.026}_{-0.021}$ & 0.168$\pm{0.032}$ \\ % par1
$kT_1$  (keV)                               	& 0.166$\pm{0.002}$ & 0.165$\pm{0.008}$ & 0.174$^{+0.006}_{-0.008}$ \\ % par2
$kT_2$  (keV)                                	& 0.494$^{+0.022}_{-0.015}$ & 0.496$^{+0.052}_{-0.050}$  & 0.507$^{+0.060}_{-0.044}$ \\ % par18
%C       (solar)                              		& 0.3 (fixed)	&  0.3 (fixed)  & 0.3 (fixed) \\
%N       (solar)                              		& 0.3 (fixed)	&  0.3 (fixed)  & 0.3 (fixed) \\
O       (solar)                              		& 0.26$\pm{0.01}$& 0.24$^{+0.05}_{-0.03}$ & 0.23$^{+0.03}_{-0.02}$  \\ % par6
Ne      (solar)                              	& 0.55$^{+0.02}_{-0.03}$	& 0.49$^{+0.09}_{-0.07}$ & 0.48$\pm{0.06}$ \\ % par7
Mg      (solar)                              	& 0.38$^{+0.03}_{-0.04}$	& 0.36$^{+0.11}_{-0.08}$ & 0.30$^{+0.08}_{-0.07}$  \\ % par8
%Al      (solar)                              		& 0.3 (fixed)	&  0.3 (fixed)   \\
Si      (solar)                              		& 0.60$^{+0.08}_{-0.11}$	&  0.61$^{+0.29}_{-0.20}$ & 0.50$^{+0.18}_{-0.15}$ \\ % par10
S       (solar)                              		& 1.13$\pm{0.26}$ & $1.51^{+1.16}_{-0.73}$ & 0.80$^{+0.70}_{-0.58}$ \\ % par11
%Ar      (solar)                              		& 0.3 (fixed)	&  0.3 (fixed)  \\ 
%Ca      (solar)                              	& 0.3 (fixed)	&  0.3 (fixed)  \\ 
Fe      (solar)                              		& 0.32$^{+0.01}_{-0.02}$	& 0.30$^{+0.08}_{-0.05}$ & 0.25$^{+0.11}_{-0.08}$ \\ % par14
%Ni      (solar)                              		& 0.3 (fixed)	&   0.3 (fixed) \\ 
log$EM_1$ (cm$^{-3}$ arcmin$^{-2}$)		& 55.28$\pm{0.02}$	& 55.37$^{+0.11}_{-0.10}$ & 55.22$^{+0.12}_{-0.11}$\\ % par17 
log$EM_2$ (cm$^{-3}$ arcmin$^{-2}$)      		& 54.02$\pm{0.02}$	& 54.03$^{+0.09}_{-0.08}$ & 54.01$^{+0.07}_{-0.08}$ \\ % par33
%
%Flux1 in 0.2--5 keV (10$^{-14}$ erg s$^{-1}$ cm$^{-2}$ arcmin$^{-2}$)			& 2.7$\pm0.1$	&  3.7$^{+1.1}_{-0.7}$ & 2.8$^{+0.9}_{-0.7}$ \\ % 
Flux1 in 0.4--5 keV (10$^{-14}$ erg s$^{-1}$ cm$^{-2}$ arcmin$^{-2}$)			& 2.5$\pm0.1$	&  3.3$^{+1.0}_{-0.6}$ & 2.6$^{+0.8}_{-0.7}$ \\ % 
\hspace{2.5em} in 0.5--7 keV (10$^{-14}$ erg s$^{-1}$ cm$^{-2}$ arcmin$^{-2}$)	& 2.2$\pm0.1$	&  2.9$^{+0.9}_{-0.5}$ & 2.3$^{+0.7}_{-0.6}$ \\ % 
%Flux2 in 0.2--5 keV (10$^{-14}$ erg s$^{-1}$ cm$^{-2}$ arcmin$^{-2}$)			& 0.81$\pm0.05$	&  0.91$^{+0.10}_{-0.17}$  & 0.77$^{+0.14}_{-0.13}$ \\ 
Flux2 in 0.4--5 keV (10$^{-14}$ erg s$^{-1}$ cm$^{-2}$ arcmin$^{-2}$)			& 0.80$\pm0.05$	&  0.90$^{+0.10}_{-0.17}$  & 0.76$^{+0.14}_{-0.13}$ \\
\hspace{2.5em} in 0.5--7 keV (10$^{-14}$ erg s$^{-1}$ cm$^{-2}$ arcmin$^{-2}$)	& 0.79$\pm0.05$	&  0.88$^{+0.10}_{-0.16}$  & 0.75$^{+0.14}_{-0.13}$ \\
%
%Luminosity1 in 0.2--5 keV (10$^{34}$ erg s$^{-1}$)	     	     & 6.40$\pm$0.24 & 1.55$^{+0.46}_{-0.29}$ & 1.94$^{+0.62}_{-0.48}$ \\
Luminosity1 in 0.4--5 keV (10$^{34}$ erg s$^{-1}$)	   	     & 2.03$\pm$0.08 & 0.43$^{+0.17}_{-0.08}$ & 0.72$^{+0.23}_{-0.18}$ \\
\hspace{5.25em} in 0.5--7 keV (10$^{34}$ erg s$^{-1}$)	     & 1.69$\pm$0.07 & 0.35$^{+0.10}_{-0.07}$ & 0.60$^{+0.19}_{-0.15}$ \\
%Luminosity2 in 0.2--5 keV (10$^{34}$ erg s$^{-1}$)              & 0.41$\pm$0.03 & 0.079$^{+0.008}_{-0.014}$ & 0.15$\pm{0.03}$ \\
Luminosity2 in 0.4--5 keV (10$^{34}$ erg s$^{-1}$)         	     & 0.34$\pm$0.02 & 0.064$^{+0.006}_{-0.011}$ & 0.12$\pm{0.02}$ \\
\hspace{5.25em} in 0.5--7 keV (10$^{34}$ erg s$^{-1}$)         & 0.32$\pm$0.02 & 0.060$^{+0.006}_{-0.011}$ & 0.11 $\pm{0.02}$ \\
\hline									   									   
								   								   
Power-law component$^{\rm c}$        \\  				   				   
%Flux in 0.2--5 keV (10$^{-14}$ erg s$^{-1}$ cm$^{-2}$ arcmin$^{-2}$)    & 0.54$^{+0.02}_{-0.01}$	& 0.49$^{+0.04}_{-0.03}$ & 0.55$\pm{0.03}$ \\% par35
Flux in 0.4--5 keV (10$^{-14}$ erg s$^{-1}$ cm$^{-2}$ arcmin$^{-2}$)       & 0.54$^{+0.02}_{-0.01}$	& 0.49$^{+0.04}_{-0.03}$ & 0.55$\pm{0.03}$ \\% par35
\hline									   									   
									   									   
LHB component$^{\rm d}$              \\					   					   
%Flux in 0.2--5 keV (10$^{-14}$ erg s$^{-1}$ cm$^{-2}$ arcmin$^{-2}$)	& 0.21$\pm0.04$ & 0.21 (fixed) & 0.17$\pm{0.06}$\\ % par39
Flux in 0.4--5 keV (10$^{-14}$ erg s$^{-1}$ cm$^{-2}$ arcmin$^{-2}$)	& 0.05$\pm0.01$ & 0.05 (fixed) & 0.04$\pm{0.02}$\\ % par39
\hline									   									   
									   									   
$\chi^2$/d.o.f.			&  1.23  & 1.09 & 1.17	\\
         d.o.f.			&  723 & 255 & 427	\\

\hline

    \end{tabular}

  \end{center}
{\noindent

  $^{\rm a}$ Fitting models.  Fitting errors at the 90\% confidence level are shown. The luminosity is corrected for absorption. \\
  $^{\rm b}$ {\tt phabs * (apec $+$ apec $+$ power-law) $+$ apec} (LHB).
Abundances of elements that are not shown in this table are fixed at 0.3 solar values. 
The absolute solar abundance tables are referred to \citet{and89}.
  $^{\rm c}$ The photon index is fixed at 1.5. The same
  absorption for the two-temperature plasma is assumed. \\
%  Normalization is photon keV$^{-1}$ cm$^{-2}$ at 1 keV.\\
  $^{\rm d}$ The single-temperature plasma model representing LHB.
  A plasma temperature is fixed at 0.1 keV.
  No absorption is assumed for this foreground component. \\
}
\end{table}

\subsection{Multiple absorption models}

\begin{figure}[htbp]
	\centerline{
		\includegraphics[width=0.8\textwidth]{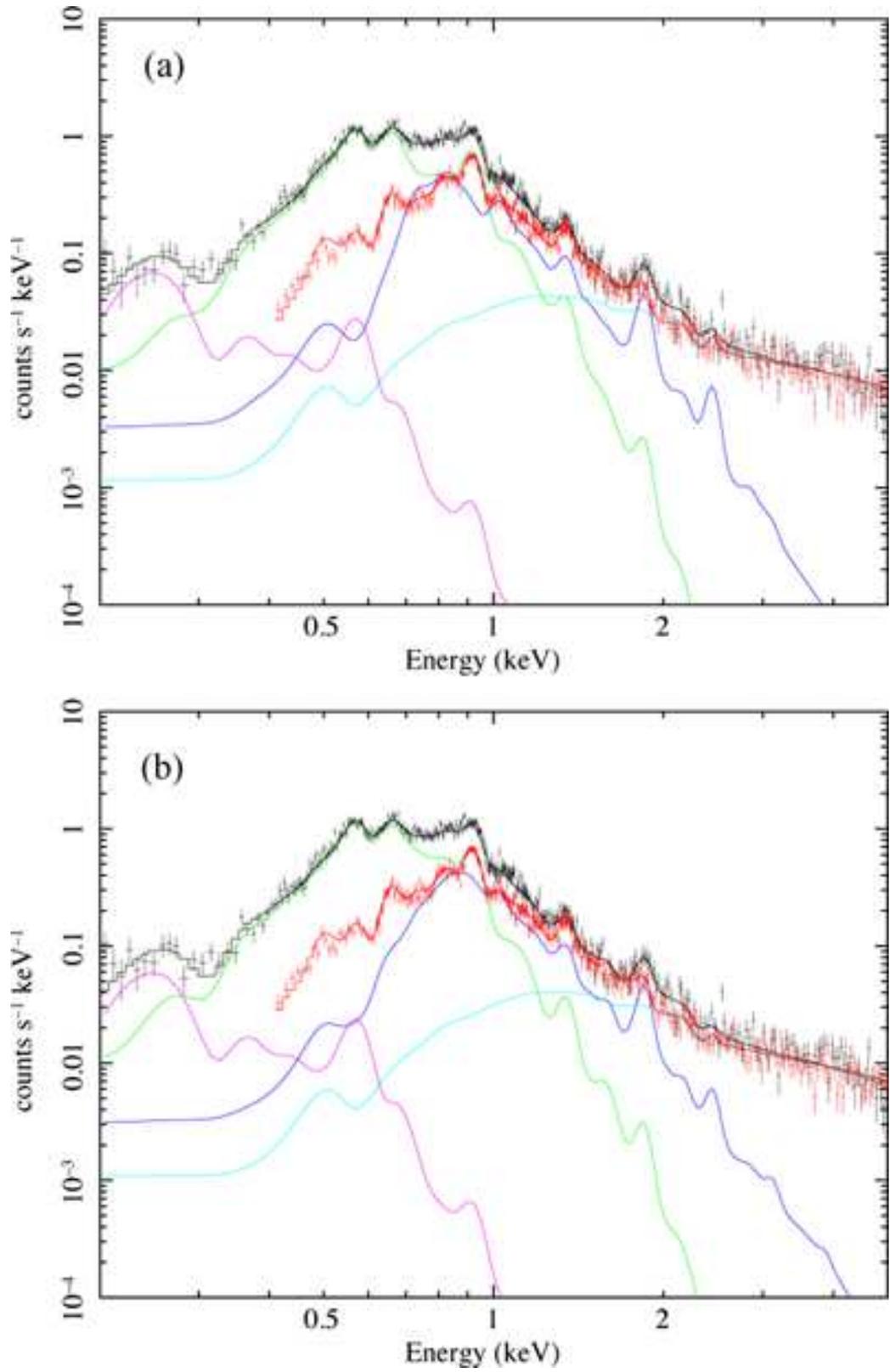}
	}
	\smallskip
	\caption{The same as figure \ref{fig5} but for two-temperature (a) CIE and (b) NEI plasma model with different absorptions. See table \ref{tbl:fit2} for the obtained parameters.}
\label{fig6}
\end{figure}

\citet{tow11b} adopted a multi-temperature non-equilibrium (NEI)  plasma model 
components, each of which suffer individual absorption.
With the high photon statistics of the {\it Suzaku} XIS spectra, we test 
these possibilities. We try a two-temperature collisional ionization (CIE) plasma model
assuming different absorption for the individual temperature component. The result is shown 
in figure \ref{fig6} (a) and table \ref{tbl:fit2}. 
The reduced chi-square value is not significantly better than that of the common absorption model.
The absorption column of the hot component is not well constrained while the other parameters are
similar to those of the common absorption model. 

We also try a two-temperature NEI plasma components suffering different absorptions. 
We adopt {\tt vpshock}, following to \citet{tow11b}.
The result is shown in figure \ref{fig6} (b) and table \ref{tbl:fit2}. 
The reduced chi-square value is is slightly worse than that of the previous models.
The upper limit of $\tau_{\rm u}$ is near the CIE timescale of $\sim3\times10^{12}$ s 
cm$^{-3}$, as expected from the acceptable fittings of the CIE models. 

These multiple absorption models therefore do not significantly improve the spectral fits.  
Below we employ the two-temperature CIE plasma model with common absorption for 
further analysis and discussion.

%%% Table 

\begin{table}[p]
  \caption{Multiple absorption model fitting results for spectra of the whole region.}

  \label{tbl:fit2}

  \begin{center}
    \begin{tabular}{lccccc}

\hline
Model		                                 &  CIE$^{\rm a}$ & NEI$^{\rm b}$ \\
\hline									       	        								   

$N_{\rm H1}$ (10$^{22}$ cm$^{-2}$)	& 0.181$^{+0.005}_{-0.008}$ & 0.125$\pm{0.017}$ \\
$kT_1$  (keV)                               	& 0.169$\pm{0.003}$ & 0.213$^{+0.007}_{-0.014}$  \\
$N_{\rm H2}$ (10$^{22}$ cm$^{-2}$)	& 0.253$^{+0.018}_{-0.022}$ & 0.275$^{+0.110}_{-0.105}$ \\ 
$kT_2$  (keV)                                	& 0.486$^{+0.082}_{-0.013}$ & 0.681$^{+0.065}_{-0.059}$ \\
O       (solar)                              		& 0.25$^{+0.02}_{-0.01}$& 0.22$\pm{0.01}$  \\ 
Ne      (solar)                              	& 0.53$^{+0.03}_{-0.02}$	& 0.41$\pm{0.03}$  \\
Mg      (solar)                              	& 0.37$^{+0.04}_{-0.03}$	& 0.26$\pm{+0.03}$   \\ 
Si      (solar)                              		& 0.54$^{+0.07}_{-0.09}$	&  0.28$\pm{0.06}$ \\ 
S       (solar)                              		& 0.95$\pm{0.23}$ & 0.26$^{+0.18}_{-0.17}$ \\
Fe      (solar)                              		& 0.38$^{+0.02}_{-0.05}$	& 0.24$^{+0.04}_{-0.03}$ \\
$\tau_{\rm l}$ (10$^{10}$ s cm$^{-3}$)              & -- & 1.0 (fixed) \\
$\tau_{\rm u}$ (10$^{11}$ s cm$^{-3}$)              & -- & 8.0$^{+3.2}_{-2.8}$\\
log$EM_1$ (cm$^{-3}$ arcmin$^{-2}$)		& 55.25$^{+0.02}_{-0.01}$	& 54.87$\pm{0.08}$ \\
log$EM_2$ (cm$^{-3}$ arcmin$^{-2}$)      		& 54.06$\pm{0.03}$	& 54.05$^{+0.09}_{-0.15}$ \\
%Flux1 (10$^{-14}$ erg s$^{-1}$ cm$^{-2}$ arcmin$^{-2}$)					& 2.8$\pm0.1$	&  2.9$^{+0.6}_{-0.5}$  \\
Flux1 in 0.4--5.0 keV (10$^{-14}$ erg s$^{-1}$ cm$^{-2}$ arcmin$^{-2}$)			& 2.6$\pm0.1$	&  2.6$\pm{0.5}$\\
\hspace{2.5 em} in 0.5--7.0 keV (10$^{-14}$ erg s$^{-1}$ cm$^{-2}$ arcmin$^{-2}$)	& 2.3$\pm0.1$	&  2.4$^{+0.5}_{-0.4}$  \\
%Flux2 (10$^{-14}$ erg s$^{-1}$ cm$^{-2}$ arcmin$^{-2}$)					& 0.77$^{+0.06}_{-0.05}$	&  0.74$^{+0.16}_{-0.22}$  \\
Flux2 in 0.4--5.0 keV (10$^{-14}$ erg s$^{-1}$ cm$^{-2}$ arcmin$^{-2}$)			& 0.77$^{+0.06}_{-0.05}$	&  0.74$^{+0.16}_{-0.22}$  \\
\hspace{2.5 em} in 0.5--7.0 keV (10$^{-14}$ erg s$^{-1}$ cm$^{-2}$ arcmin$^{-2}$)	& 0.76$^{+0.06}_{-0.05}$	&  0.74$^{+0.16}_{-0.22}$  \\
%Luminosity1 (10$^{34}$ erg s$^{-1}$)					 & 5.79$\pm{0.21}$ & 2.23$^{+0.47}_{-0.40}$ \\
Luminosity1 in 0.4--5.0 keV (10$^{34}$ erg s$^{-1}$)			 & 1.94$\pm{0.07}$ & 1.38$^{+0.29}_{-0.25}$ \\
\hspace{5.25em} in 0.5--7.0 keV (10$^{34}$ erg s$^{-1}$) 		 & 1.61$\pm{0.06}$ & 1.10$^{+0.23}_{-0.19}$ \\
%Luminosity2 (10$^{34}$ erg s$^{-1}$)					 & 0.49$^{+0.04}_{-0.03}$ & 0.47$^{+0.10}_{-0.14}$\\
Luminosity2 in 0.4--5.0 keV (10$^{34}$ erg s$^{-1}$) 		& 0.40$^{+0.03}_{-0.02}$ & 0.41$^{+0.09}_{-0.10}$\\
\hspace{5.25em} in 0.5--7.0 keV (10$^{34}$ erg s$^{-1}$) 		& 0.38$^{+0.03}_{-0.02}$ & 0.39$^{+0.08}_{-0.10}$\\
\hline									   									   
								   								   
Power-law component$^{\rm c}$        \\  				   				   
%Flux in 0.2--5.0 keV (10$^{-14}$ erg s$^{-1}$ cm$^{-2}$ arcmin$^{-2}$)    & 0.52$^{+0.02}_{-0.01}$	& 0.50$^{+0.01}_{-0.03}$ \\
Flux in 0.4--5.0 keV (10$^{-14}$ erg s$^{-1}$ cm$^{-2}$ arcmin$^{-2}$)       & 0.52$^{+0.02}_{-0.01}$	& 0.49$^{+0.01}_{-0.03}$ \\
\hline									   									   
									   									   
LHB component$^{\rm d}$              \\					   					   
%Flux in 0.2--5.0 keV (10$^{-14}$ erg s$^{-1}$ cm$^{-2}$ arcmin$^{-2}$)	& 0.21$\pm0.04$ & 0.18$^{+0.05}_{-0.03}$ \\
Flux in 0.4--5.0 keV (10$^{-14}$ erg s$^{-1}$ cm$^{-2}$ arcmin$^{-2}$)	& 0.04$\pm0.01$ & 0.04$\pm{0.01}$ \\
\hline									   									   
									   									   
$\chi^2$/d.o.f.			&  1.23  & 1.25 \\
         d.o.f.			&  722 & 721 \\
\hline

    \end{tabular}

  \end{center}
{\noindent

%  $^{\rm a}$ Fitting models. Fitting errors at the 90\% confidence level are shown. The luminosity is corrected for absorption. \\ 
  $^{\rm a}$ Two-temperature CIE plasma model with a multiple absorption model: {\tt phabs * apec $+$ phabs * apec $+$ phabs * power-law $+$ raymond}. 
  $^{\rm b}$ Two-temperature NEI plasma model with a multiple absorption model: {\tt phabs * vpshock $+$ phabs * vpshock $+$ phabs * power-law $+$ raymond}. $\tau_{\rm l}$ and $\tau_{\rm u}$ are lower and upper limits on ionization timescale, respectively.
  $^{\rm c}$ The photon index is fixed at 1.5. In the multi absorption model, the larger absorption for the two-temperature plasma, i.e., $N_{\rm H2}$, is assumed for this background component. 
  $^{\rm d}$ The single-temperature plasma model representing LHB.  A plasma temperature is fixed at 0.1 keV.  No absorption is assumed for this foreground component. 
}
\end{table}

\begin{figure}[htbp]
	\centerline{
		\includegraphics[width=0.7\textwidth]{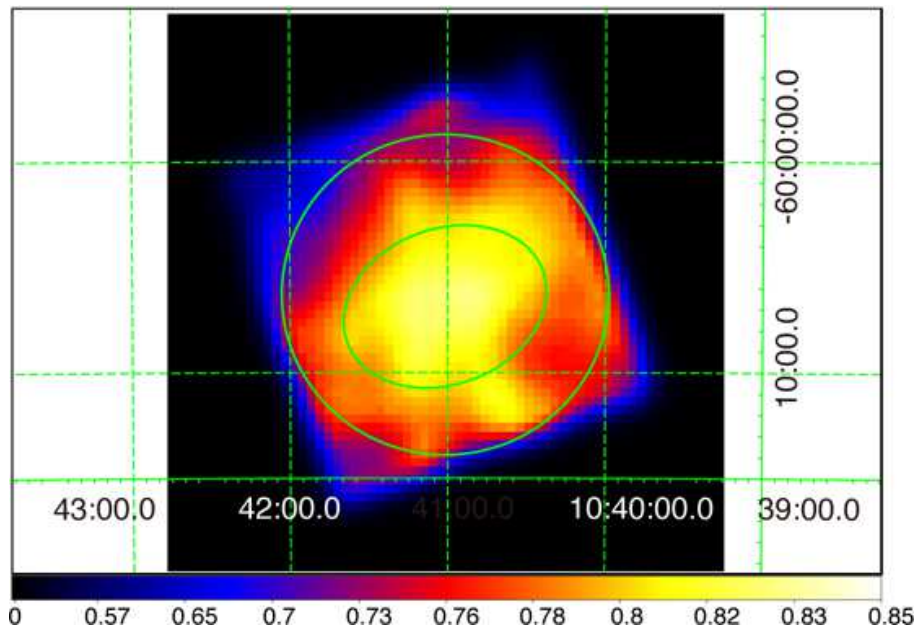}
	}
	\smallskip
	\caption{Hardness ratio map by calculating $(S-H)/(S+H)$ where $H$ and $S$ 
	are 0.8--2.0 keV and 0.4--0.8 keV XIS1 images. Before division, the 
	NXB is subtracted from each image binned by 3 pixels for limited
	photon statistics. For clarity, the hardness ratio map is smoothed by 
	a 2-dimensional Gaussian function with $\sigma=5$ pixels. 
	Solid lines indicate two subdivided regions.}
\label{fig7}
\end{figure}

\begin{figure}[htbp]
	\centerline{
		\includegraphics[width=0.8\textwidth]{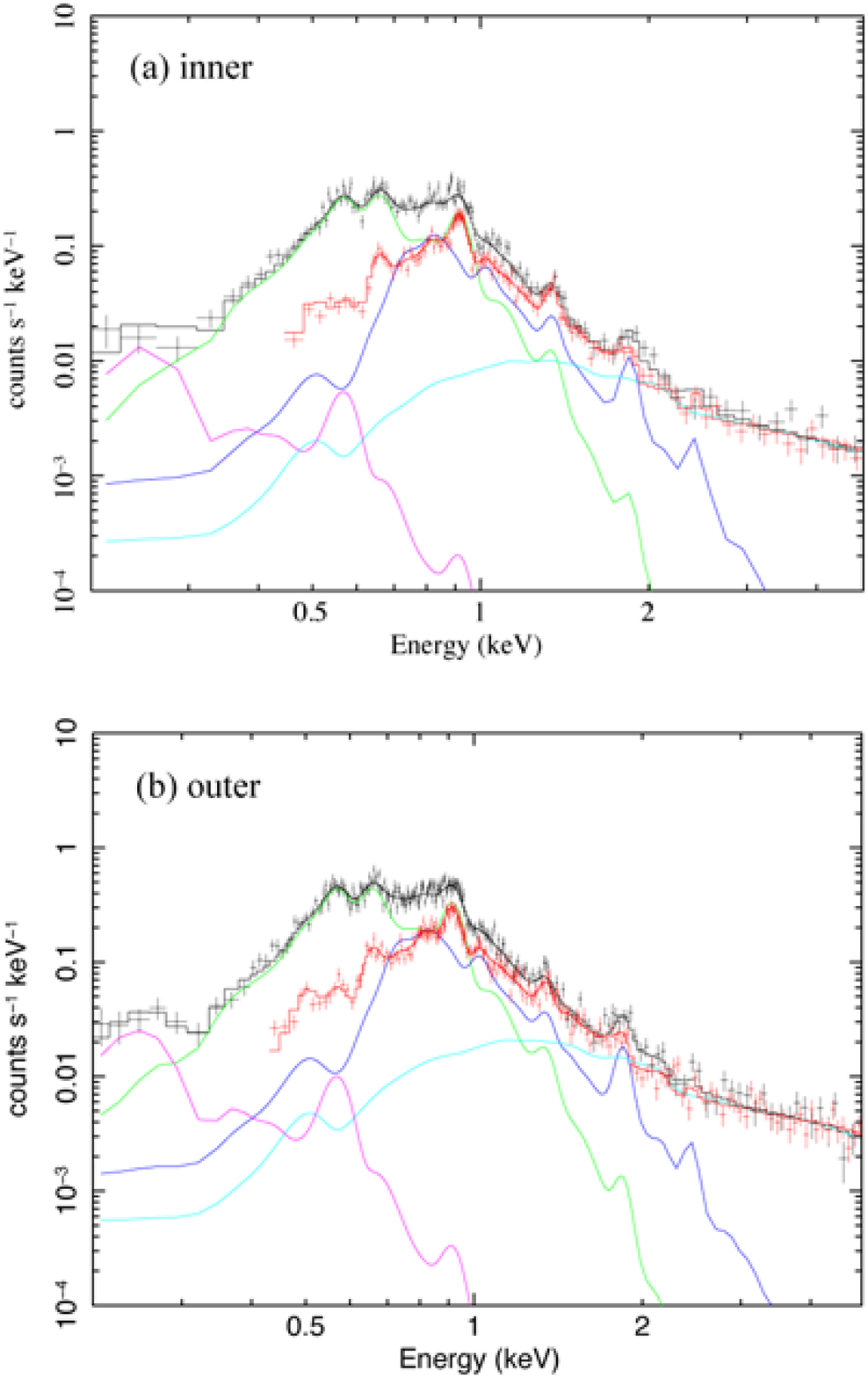}
	}
	\smallskip
	\caption{The same as figure \ref{fig5} but for (a) the inner and outer regions. See table \ref{tbl:fit1} for the obtained parameters.}
\label{fig8}
\end{figure}

\subsection{Subdivided regions}

We then investigate spatial variation of the diffuse spectrum.
We create a map of of the hardness ratio defined from images that correct the spatial effective area variation
due to the mirror vignetting and OBF contamination\footnote{Suzaku Data Reduction Guide : https://heasarc.gsfc.nasa.gov/docs/suzaku/analysis/abc/}
The map in figure \ref{fig7} shows significant softness toward the FOV center where the 
red blob is. In fact, the blob is clearly seen in the 0.4--0.8 keV image but not in the 0.8--2 keV 
image. This result suggests that the blob originates from the spatial structure of the cool 
component.

We therefore define two regions in the XIS FOV based on this hardness ratio map; the inner 
region with the softest color and the outer region that is the outskirt. We extract their spectra 
and fit them using the two-temperature CIE plasma model with common absorption.
The sizes of the inner and outer regions are 57 and 127 arcmin$^2$ for each, which correspond 
to 25 and 57 pc$^2$ at a distance of 2.3 kpc, respectively.

The applied model reproduces the two spectra well (figure \ref{fig8} and table \ref{tbl:fit1}).
Their best-fit parameters are very similar to those of the whole region within 90 \% statistical uncertainties.
We fix the flux of the LHB component for the inner region at the best fit value of the whole region 
because it became unreasonably low (0.6$\times10^{-14}$ erg s$^{-1}$ cm$^{-2}$ arcmin$^{-2}$) if it
is a free parameter probably due to the limited statistics below 0.4 keV. Since the spatial variation of the 
LHB component on arcmin scales is not likely, we fix this parameter. The obtained reduced chi-square is 
acceptable. 

The total flux of the point sources detected with {\it Chandra} in the inner and outer regions is 
$\sim$5$\times10^{-14}$ and $\sim$3$\times10^{-13}$
erg s$^{-1}$ cm$^2$ at 0.4--5 keV, whose contribution is $\sim$2 \% and $\sim$6 \% for 
the diffuse X-ray emission in the inner and outer regions, respectively. 
Therefore, similar to the whole region, the diffuse X-ray emission is dominant.

The inner region has a $\sim$30\% larger flux of the cool component than the outer region, while there was 
$\sim$20 \% difference in the flux of the hot component. This result may confirms that the cool component 
produces the blob-like structure. 

To check possible variation of absorption column density, we apply the best fit 
model of the inner region to the outer region spectra while a scaling factor and an additional 
absorption for the two temperature plasma component are free. The fitting is acceptable 
with $\chi^2$/d.o.f. of 1.20, the scaling factor of 1.01$\pm0.02$ and the additional $N_{\rm H}$
of $(0.037\pm0.004)\times10^{22}$ cm$^{-2}$. This correspond to optical extinction $A_V$ of 
$\sim$0.2 \citep{guv09}, which is small and not suggested in the past CO and optical observations.
Therefore, if there is a variation of the absorption column density, the variation is small. We thus 
simply employ the commonly absorbed two-temperature plasma model for further discussion below.

%%%%%%%%%%%%
\section{Discussion}
%%%%%%%%%%%%

We here compare this result with the previous {\it Suzaku} studies.
The three Suzaku observations were performed at different periods
--- the $\eta$ Car region was observed in 2005 August, the eastern-tip region 2006 June, 
and the Carina SW region 2010 December.
An important caveat is that their soft band spectra cannot be directly compared 
as the XIS's soft band sensitivity significantly declined between these observations 
due to contamination on the XIS blocking filter.
This sensitivity degradation is considered in {\tt arf} for spectral fitting \citep{ish07},
so we compare the results of their spectral fits, 
while we cannot directly compare their spectra in count rates due to this difference in sensitivity.

%%%%%%%%%%%%
\subsection{Spatial Distribution of the Diffuse Plasma}
%%%%%%%%%%%%

In all studies, the diffuse spectra were reproduced by models with the same components: 
two temperature thermal plasma emission suffering weak absorption.
The only difference in the fitting conditions is that 
the diffuse spectra around $\eta$ Carina were fitted with a free parameter for the carbon elemental abundance \citep{ham07}.
Without any clear carbon emission lines between 0.3$-$0.4 keV,
the fit deduced small carbon abundance.
A carbon does not provide enough free electrons to the plasma.
The hydrogen (and helium) elemental abundance are increased to supply electrons  
for Bremsstrahlung continuum emission.
This resulted in moderately low elemental abundance values to the other heavy elements.
However, X-ray emission from the Carina nebula suffers strong cut-off below 0.4 keV,
and the spectra below 0.4 keV should be strongly contaminated by foreground emission 
and/or affected by structures in the instrumental response. 
Fortunately, this conditional difference does not significantly affect measurements of the 
key spectral parameters ($N_{\rm H}$, $kT$, $EM$) 
nor abundance ratios between heavily elements.
We thus use the best-fit parameters in table 1 individual models of \citet{ham07} for the $\eta$ Car north and south regions, 
and table 3 model 1 of \citet{ezo09} for the eastern tip region for comparison.

\begin{figure}[htbp]
	\centerline{
		\includegraphics[width=0.9\textwidth]{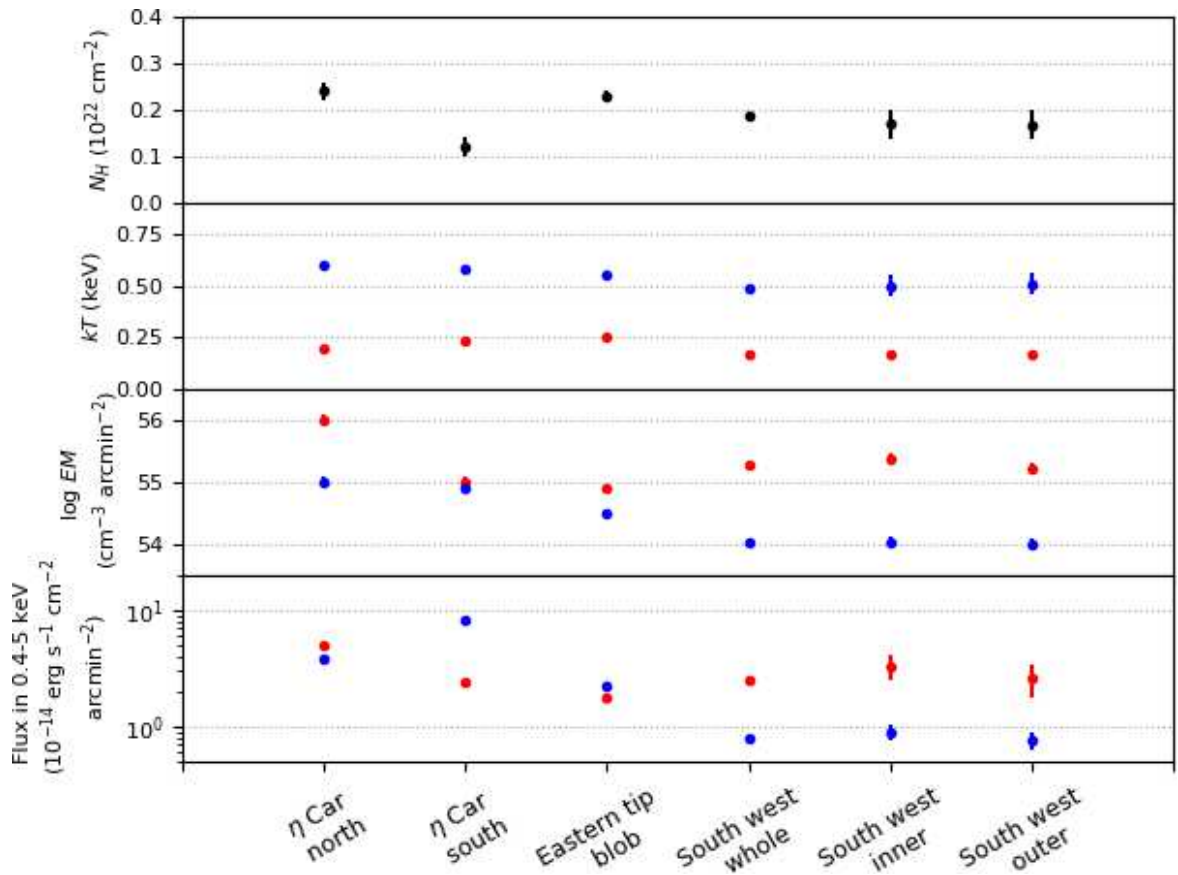}
	}
	\smallskip
	\caption{Comparison of the two-temperature plasma model fits ($\eta$Car north/south: table 1 individual models 
	in \citet{ham07}, the eastern tip blob region: table 3 model 1 of \cite{ezo09}, South west: this work). Blue and red colors 
	represent the hot and cool components.}
\label{fig9}
\end{figure}

Figure~\ref{fig9} plots $N_{\rm H}$, $kT$, $EM$ and surface brightness of the hot and cool plasma components.
The $N_{\rm H}$ values are similar between the regions.
This means that diffuse emission from these regions does not suffer significant local absorption. 
Even if we adopt the multiple absorption CIE or NEI plasma models, the larger $N_{\rm H}$ value is still small ($\sim$0.3$\times10^{22}$ cm$^{-2}$).
This is not surprising for the Carina SW and the eastern tip regions with little molecular gas in 
their lines of sight \citep{yon05}.
However, both the north and south $\eta$ Car regions have moderate amount of molecular gas columns.  
This result suggests that i) the molecular gases are not thick enough for the X-ray emission,
ii) the molecular gases are much smaller than the spatial resolution of the radio telescope (2.7$'$)
so that molecular gas does not obscure most X-ray emission,  
or iii) the molecular gases are behind the X-ray plasma.
In either case, absorption does not cause the redness of the Carina SW region.

Both cool and hot plasmas have similar temperatures between the regions;
only the Carina SW region has a slightly lower temperature by 0.05$-$0.1~keV in each component
than the other regions.
The surface brightness of the cool plasma emission lies within a factor of $\sim$2$-$3 between the regions,
while that of the hot plasma emission varies by an order of magnitude.
The Carina SW region has the lowest surface flux of the hot plasma component, and 
the lowest relative flux over the cool plasma emission.
This is the reason why the Carina SW region is colored in red.

These results suggest that there are physically two distinctive plasma components
in the Carina nebula.
The plasma temperatures of the Carina SW region are slightly cooler, possibly due to 
weaker heating and/or stronger cooling near a nebula edge.

%%%%%%%%%%%%
\subsection{Plasma Elemental Abundances}
%%%%%%%%%%%%

The spectral fits constrain elemental abundances of O, Ne, Mg, Si, S and Fe from 
their K or L-shell emission lines seen between 0.4$-$3 keV.
The Si and S abundance measurements rely mostly on their K-shell emission lines at $\sim$2 and 2.6 keV,
but background emission is comparable to the diffuse emission in this energy range, and some contamination 
sources such as GRXE and low mass young stars emit those lines as well. 
We therefore calculate abundance ratios of O, Ne and Mg relative to Fe and compare them with those
expected from SNe and stellar wind bubbles.

\begin{figure}[htbp]
	\centerline{
		\includegraphics[width=0.9\textwidth]{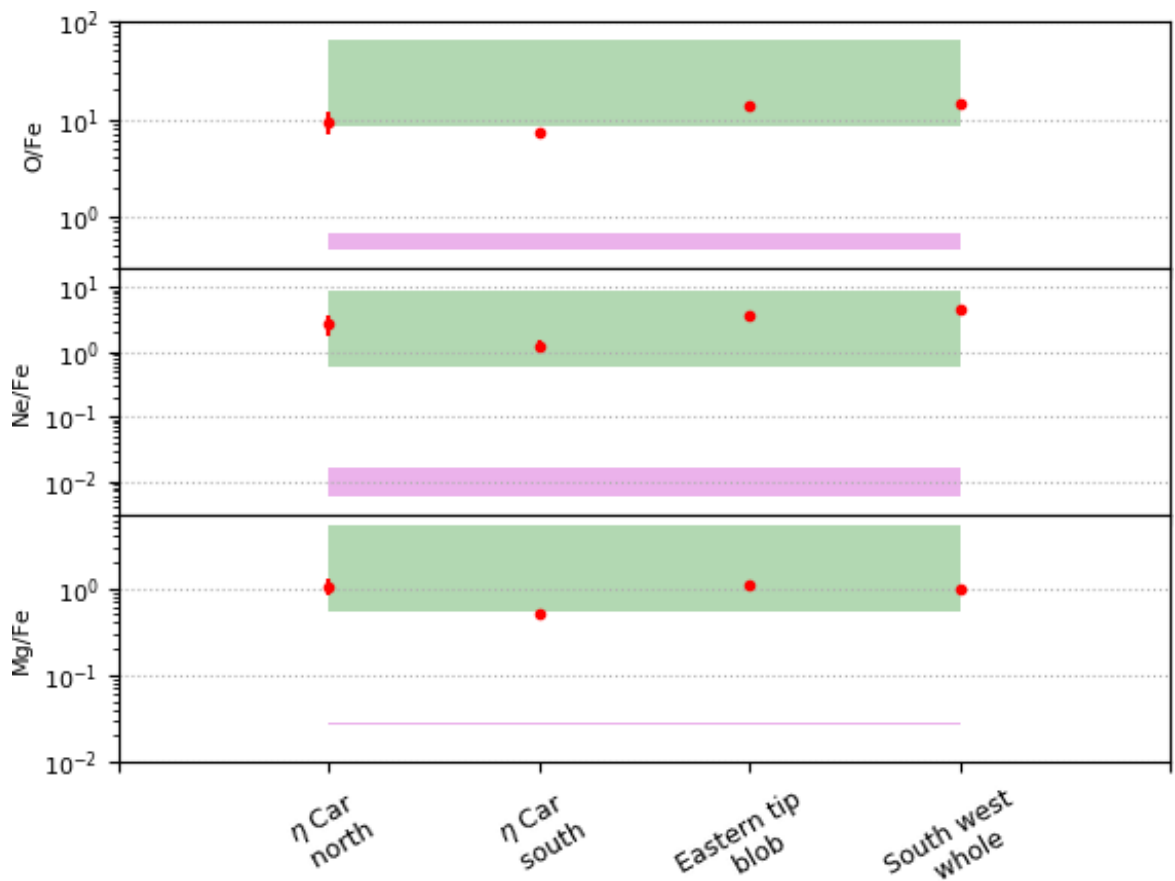}
	}
	\smallskip
	\caption{Comparison of the elemental abundance ratios. The filled green box indicates a range of the element abundance ratio 
		based on nucleosynthesis models for Type II SNe with 15--20 $M_\odot$ progenitor masses \citep{nom97},
		while the filled magenta box is that for two Type Ia SNe models
		(W7$=$fast deflagration model and WDD1$=$slow deflagration, delayed detonation model) \citep{iwa99}.}
\label{fig11}
\end{figure}

\begin{figure}[htbp]
	\centerline{
		\includegraphics[width=0.8\textwidth]{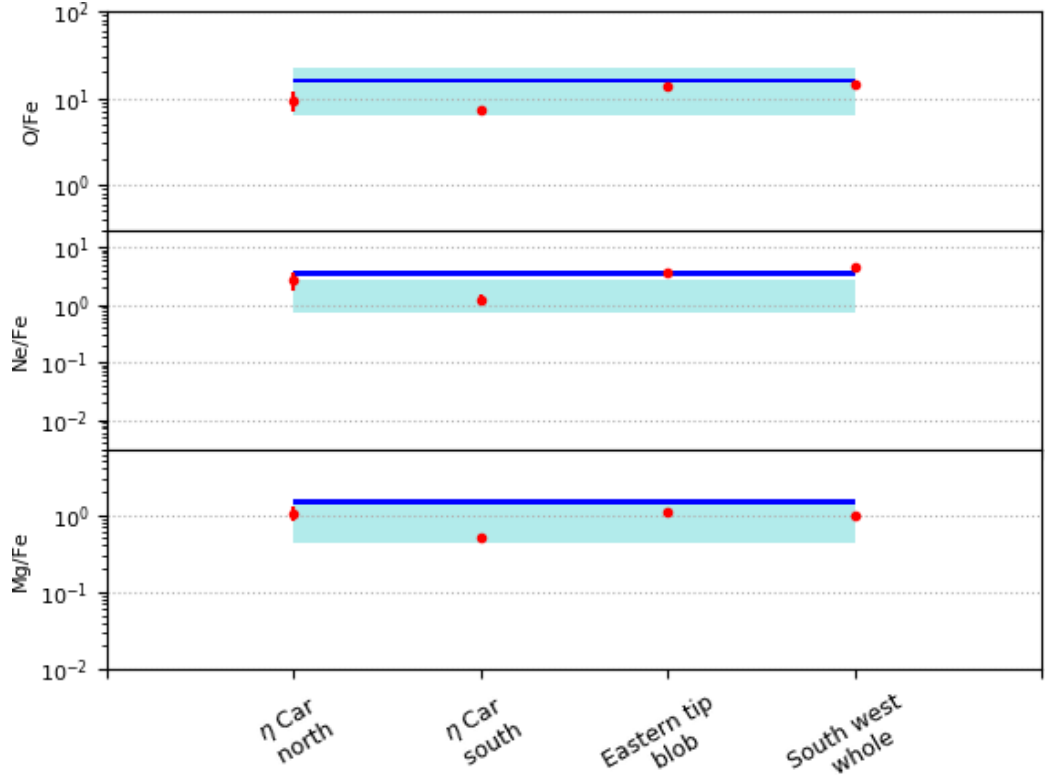}
	}
	\smallskip
	\caption{The same as figure \ref{fig9} but with the ranges of $\delta$Ori (O9.5II, \cite{ra13}) and $\tau$Sco (B0.2V, \cite{mew03}). 
	The filled blue and cyan boxes indicate 1$\sigma$ confidence regions for $\delta$Ori and  $\tau$Sco, respectively.}
\label{fig12}
\end{figure}

Figures~\ref{fig11} and \ref{fig12} plot abundance ratios of all Carina regions. 
Because the fitting errors of the elemental abundances of the inner and outer regions are relatively large and their values are consistent with those of the whole regions within errors, we plot only the whole region for the Carina SW.
Figure~\ref{fig11} also displays abundance ratios of Type Ia and core-collapsed SN remnants 
derived from theoretical models (e.g., \cite{hug95, hen03, wil05}).
The measured abundance ratios of the Carina diffuse nebula fall within 7.3$-$14.8 for O/Fe, 1.2$-$4.5 for Ne/Fe, and 0.5$-$1.1 for Mg/Fe.
These values are consistent with those of core collapsed SNe models but significantly higher than the Type Ia SNe model, which are rich in iron.

On the other hand, figure~\ref{fig12} also displays the abundance ratio of wind driven X-ray emission from OB stars; 
$\delta$Ori \citep{ra13} and $\tau$Sco \citep{mew03}.
A caveat is that the $\tau$Sco measurements are apparently worse 
than the $\delta$Ori measurements, but this is because the Fe abundance was measured independently for $\tau$Sco
(0.9$\pm$0.5 solar photospheric value, see table 2 in \cite{mew03}). 
All ratios are similar to those of OB stars.

%%%%%%%%%%%%
\subsection{Physical properties of the diffuse plasma}
%%%%%%%%%%%%

We estimate physical properties of the diffuse plasma in the Carina SW region using the same 
method as \citet{ezo09}. Here we need to assume a shape of the diffuse plasma to estimate its
volume. Even though the size of the diffuse plasma in the line of sight direction is unclear, we here simply assume 
10 pc from the size of the nebula. A side length of the XIS FOV of 17.8 arcmin corresponds to 11.9 pc at 
the distance of 2.3 kpc.
For each plasma component,
we deduce the electron density from the emission measure and the assumed plasma volume
and the pressure from the plasma temperature and the electron density. 
We then calculate the total plasma energy, its cooling time, and the mass using these physical values
as follows. 
\begin{eqnarray}
L_{\rm X} & = & \Lambda(T) EM  = \Lambda(T) n_{\rm e}^2 V \eta, \\
U & = & 3~ n_{\rm e} kT V, \\
p & = & 2~ n_{\rm e} kT, \\
t_{\rm cool} & = & U/L_{\rm X}, \\
M_{\rm plasma} & = & \mu_{\rm X}~ m_{\rm p} n_{\rm e} V,
\end{eqnarray}
where $\Lambda$, $EM$, $\eta$, and $\mu_{\rm X}$ 
denote the cooling function, the volume emissivity, a scale factor of the volume of the plasma,
and the mean molecular weight per electron. 
%%We approximate the cooling function of a low density
%%plasma of a solar metalicity as 1.0$\times10^{-23} \cdot (kT)^{0.5}$  erg s$^{-1}$ cm$^3$ s$^{-1}$.
We assumed $\Lambda =$15$\times10^{-23}$ and 5$\times10^{-23}$ erg s$^{-1}$ cm$^3$ s$^{-1}$ for the low
and high temperature plasma components and $\mu_{\rm X}=$0.62 as used in \citet{tow03} and \citet{ezo09}.

Table \ref{tbl:plasma} shows the result.
The estimated plasma density and the pressure of the high and low 
temperature components are about $\times$5-10 and $\times$10-25 times
lower than those in the blob of the easter tip region (table 6 in \cite{ezo09}). 
Although these pressures can be lower than those of the plasma near $\eta$Car and 
the eastern-tip region, they are still comparable to those of molecular clouds. Since no 
strong CO emission is observed in this region \citep{yon05} (see figure \ref{fig1}), 
the plasma pressures of the two temperature components are still larger than 
ambient materials and can continue to expand.

Another important point is that the cooling times are longer than a typical life time of 
OB stars (0.1-1 Myr) and also a propagation time of the plasma at sound velocity from
 the center of this nebula (i.e., near $\eta$Car) to this area (on the order of 0.1 Myr).
Therefore, this may support the scenario that plasma generated around the center of the
Carina nebula via stellar winds of OB stars or SNe propagates toward the surrounding 
region but the plasma pressure slightly drops due to expansion. 

\begin{table}[htbp]
  \caption{Physical properties of the diffuse plasma in the whole region$^a$.}
  \label{tbl:plasma}

  \begin{center}
    \begin{tabular}{lccccc}

\hline
Parameter                                    &  Scale Factor                &       $T_1$                     &      $T_2$\\
\hline

\multicolumn{4}{c}{Observed X-ray Properties}\\
\hline

$kT$ (keV)                                   &   $-$                               &        0.2                       &      0.5 \\
$L_{\rm X}$ (ergs s$^{-1}$)     	  &   $-$                               &   2$\times10^{34}$     &     3$\times10^{33}$\\
$V$ (cm$^{3}$)                            &   $\eta$                          &   4$\times10^{58}$     &     4$\times10^{58}$\\
% print (11.9 * 3.085e18)**2 * (10 * 3.085e18)
\hline

\multicolumn{4}{c}{Estimated X-ray Plasma Properties}\\
\hline

% cooling 15e-23 erg/sec/cc @ 0.2 keV = 2.3e6 K
%                  5e-23 erg/sec/cc @ 0.6 keV = 7.0e6 K

$n_{\rm e}$ (cm$^{-3}$)             &   $\eta^{-1/2}$            &        0.06                          &      0.04 \\
% print (2.03e34/15e-23/4.16e58)**0.5
% print (3.38e33/5e-23/4.16e58)**0.5
$P/k$ (K cm$^{-3}$)                    &   $\eta^{-1/2}$            &   2$\times10^{5}$       &     2$\times10^{5}$\\
% print 2 * (2.03e34/15e-23/4.16e58)**0.5 * (0.17e3*1.60e-19/1.38e-23)/1e6
% print 2 * (3.38e33/15e-23/4.16e58)**0.5 * (0.49e3*1.60e-19/1.38e-23)/1e6
$U$ (ergs)                                     &   $\eta^{1/2}$             &   2$\times10^{48}$     &     2$\times10^{48}$\\
% print 3 * (2.03e34/15e-23/4.16e58)**0.5 * (0.17e3*1.60e-12) * 4.16e58
% print 3 * (3.38e33/15e-23/4.16e58)**0.5 * (0.49e3*1.60e-12) * 4.16e58
$t_{\rm cool}$ (yr)                         &  $\eta^{1/2}$             &   3$\times10^{6}$       &     2$\times10^{7}$\\
% print 3 * (2.03e34/15e-23/4.16e58)**0.5 * (0.17e3*1.60e-12) * 4.16e58 / (2.03e34 * 24 * 3600 * 365) /1e6
% print 3 * (3.38e33/15e-23/4.16e58)**0.5 * (0.49e3*1.60e-12) * 4.16e58 / (3.38e33 * 24 * 3600 * 365) /1e6
$M_{\rm plasma}$ ($M_\odot$) &  $\eta^{1/2}$             &   1.2                              &     0.5 \\
% print 0.62 * 1.673e-24 * (2.03e34/15e-23/4.16e58)**0.5 * 4.16e58 / 1.99e33
% print 0.62 * 1.673e-24 * (3.38e33/15e-23/4.16e58)**0.5 * 4.16e58 / 1.99e33

\hline

    \end{tabular}

  \end{center}
{\noindent

  $^{\rm a}$ $\eta$ is a scale factor for the volume of the 
             plasma. $T_1$ and $T_2$ indicate the two temperature plasma
             component in table \ref{tbl:fit1}.

}
\end{table}

%%%%%%%%%%%%
\subsection{Implications on the SNe and stellar wind bubble scenario}
%%%%%%%%%%%%

These arguments suggest that there are two diffuse plasma components in the Carina nebula:
hot ({\it kT} $\sim$0.5-0.6 keV) plasma concentrated in the center of the nebula 
and the cool ({\it kT} $\sim$0.2-0.3 keV) plasma extended throughout the nebula.
Their plasma temperatures fit with both the core-collapsed SNe and stellar wind bubble scenario. 
Relatively young ($\lesssim10^4$ yr) core-collapsed SNe have
$kT$ of 0.1--1 keV plasma in ionization equilibrium (e.g., \cite{wil05}).
Stellar wind bubbles from OB stars can also produce a shock temperature of $\sim$1 keV  with
1000 km s$^{-1}$ velocity winds \citep{ezo09}. 

\citet{tow11c} estimated the total energy of the diffuse plasma in the Carina nebula at $\sim10^{50}$ ergs.
The Carina nebula currently possesses $\sim$66 massive stars, and it is likely that the nebula would have
produced more massive stars in the past that exploded as SNe. 
Indeed, \citet{ham09} discovered a middle-aged neutron star in the Carina nebula,
whose progenitor may have erupted as a Type II SN in the nebula.

On the other hand, 
a single massive star can supply $\sim3\times10^{48}$ ergs throughout its lifetime to 
the interstellar space through wind kinematic energy (e.g., \cite{ezo09}).
Even if we assume that this energy is converted to the plasma energy 
with a certain efficiency (e.g., 0.1 \%), the total energy is only $\sim3\times10^{45}$
ergs. Therefore, we need $\sim 3\times10^4$ OB stars, which is far larger than the number
of the known OB stars in the Carina nebula ($\sim$70). Therefore, the stellar winds 
only may not explain all thermal energy of the diffuse plasma.

From these points of view, we suggest that the origin of the diffuse plasma in the Carina
nebula maybe a combination of Type II SNe and stellar winds or Type II SNe only. 
Considering the different spatial distribution over the Carina nebula,
the two temperature components may have different events/origins. 

%%%%%%%%%%%%
\section{Summary}
%%%%%%%%%%%%

We investigate diffuse X-ray emission near a south-west edge of the Carina nebula 
using the {\it Suzaku} observatory.
We carefully estimate contribution of plausible emission components in the field and conclude 
that most emission originates from truly diffuse plasma.

The diffuse X-ray spectrum is reproduced by an absorbed two-temperature CIE plasma emission 
model. 
Two-temperature CIE and NEI plasma models with multiple absorption columns can also
reproduce the data but $\chi^2$/d.o.f. values are similar. Therefore, the simple two-temperature
CIE plasma emission with single absorption is adopted.
The obtained spectral parameters in the absorbed two-temperature CIE plasma emission
model are compared to those of the other regions in the Carina nebula.
The surface brightness of the hot component is relatively lower than those of the other regions
of the Carina nebula and thus the cool plasma is evident. 
Therefore, the two temperature plasma components may have different origins.

The plasma temperatures and elemental abundances 
are consistent with the origin of Type II SNe or stellar wind bubbles,
while the total energy of the diffuse plasma in the Carina nebula may not be explained
by stellar winds from existing OB stars only.
The plasmas thus may originate from both the ancient Type II SNe and OB stellar winds
or Type II SNe only.

Future high resolution spectroscopy with X-ray microcalorimeters will provide
more detailed information of plasma properties such as plasma motions and ionization states
which should help understand the origin of diffuse X-ray emission from the Carina nebula.

\begin{ack}

We thank Leisa Townsley and Patrick Broos for providing us with the CCCP images
and the summed spectrum of the point sources. We also would like to acknowledge
the {\it Suzaku} XIS team for beautiful calibration.

\end{ack}

\end{document}